\documentclass{aa}
\usepackage{amsmath}
\usepackage{psfig}
\usepackage{astrobib}
\usepackage{array}
\usepackage{journals}

\def\be{\begin{equation}}
\def\ee{\end{equation}}
\begin{document}

\title{A Scheme to Unify Low-Power Accreting Black Holes}
\subtitle{Jet-Dominated Accretion Flows and the Radio/X-Ray
Correlation}

\author{Heino Falcke\inst{1,2,3}, Elmar K\"ording\inst{3}, Sera
  Markoff\inst{4}\fnmsep\thanks{NSF Astronomy \& Astrophysics
  Postdoctoral Fellow} }

\institute{
ASTRON, P.O. Box 2, 7990 AA Dwingeloo, The Netherlands
\and
Adjunct Professor, Astronomy Department, University of Nijmegen, Postbus 9010, 6500 GL Nijmegen, The Netherlands
\and
Max-Planck-Institut f\"ur Radioastronomie, Auf dem H\"ugel
69, D-53121 Bonn,
Germany 
\and 
Massachusetts Institute of Technology, Center for Space Research,
77 Massachusetts Ave., Rm. NE80-6035, Cambridge, MA 02139, USA
}

\titlerunning{Unifying Low-Power Black Holes}
\authorrunning{Falcke, K\"ording, Markoff}
\offprints{falcke@astron.nl}

\date{Received ; Accepted}

\abstract{We explore the evolution in power of black holes of all
  masses, and their associated jets, within the scheme of an accretion
  rate-dependent state transition.  Below a critical value of the
  accretion rate all systems are assumed to undergo a transition to a
  state where the dominant accretion mode is optically thin and
  radiatively inefficient.  In these significantly sub-Eddington
  systems, the spectral energy distribution is predicted to be
  dominated by non-thermal emission from a relativistic jet whereas
  near-Eddington black holes will be dominated instead by emission
  from the accretion disk.  Reasonable candidates for such a
  sub-Eddington state include X-ray binaries in the hard and quiescent
  states, the Galactic Center (Sgr A*), LINERs, FR~I radio galaxies,
  and a large fraction of BL~Lac objects.  Standard jet physics
  predicts non-linear scaling between the optically thick (radio) and
  optically thin (optical or X-ray) emission of these systems, which
  has been confirmed recently in X-ray binaries. We show that this
  scaling relation is also a function of black hole mass and only
  slightly of the relativistic Doppler factor. Taking the scaling into
  account we show that indeed hard and quiescent state X-ray binaries,
  LINERs, FR~I radio galaxies, and BL~Lacs can be unified and fall on
  a common radio/X-ray correlation. This suggests that jet domination
  is an important stage in the luminosity evolution of accreting black
  hole systems.   } \maketitle

\section{Introduction}
Accreting black holes are thought to be the engines powering most of
the emission from active galactic nuclei (AGN) and some X-ray binaries
(XRBs).  Associated relativistic jets also contribute significantly to
the overall spectrum, over a wide range of wavelengths.  The
current accretion paradigm is based on the early success of standard,
optically thick accretion disk models \cite{ShakuraSunyaev1973} which
correctly predicted the soft X-ray emission in stellar mass black
holes (XRBs) and the ``big blue bump'' in quasars
\cite{SandersPhinneyNeugebauer1989,SunMalkan1989} and other AGN by
scaling mass and accretion rate. Scaling laws for the radio emission
of jet cores and lobes have also been developed
\cite{FalckeBiermann1995,KaiserAlexander1997,Heinz2002} and
successfully applied to XRBs and AGN
\cite{FalckeMalkanBiermann1995,FalckeBiermann1996,FalckeBiermann1999}.

The most important parameters of accreting black holes are probably
the mass and the accretion rate, both of which can vary over many
orders of magnitude. Additional parameters which likely impact the
observable characteristics of black holes are the spin and the
inclination angle of their spin axes.  Inclination-based unified
schemes of AGN merge apparently different objects based on the angle
between the spin axis and the line of sight (see e.g.,
\citeNP{Antonucci1993,UrryPadovani1995}). 
The success of this scheme supports the evidence for angle-dependent
obscuration and relativistic beaming.

However, the exact effect that changes in the accretion rate have on
the appearance of their associated black holes systems is a matter of
ongoing debate.  A good understanding of this is crucial for modeling
the cosmic evolution of black holes and for disentangling the
different source classes.

A number of recent results suggest that the transition from a high
accretion rate black hole to a low accretion rate one is not smooth,
but rather accompanied by a ``phase transition''.  In the low-power
phase, the optically thick disk emission is either dominated by emission from
an optically thin corona, completely reduced to  a radiatively
inefficient inflow, or is truncated and an optically thin inner
radiatively inefficient flow exists closer to the compact object (see
\citeNP{Poutanen1998} for a review of the various models).  For XRBs
\citeN{EsinMcClintockNarayan1997} estimate that this transition occurs
once the accretion rate for a black hole of mas $M_\bullet$ drops to
less than a critical value ($\sim$10\% of the Eddington accretion
rate, $\dot M_{\rm Edd}\simeq2\times(M_\bullet/10^8
M_\odot)\;M_\odot\,{\rm yr}^{-1}$ for $\dot M_{\rm Edd}=L_{\rm
Edd}/0.1 c^2$).  More recent work suggests that this transition could
already occur around 2\% $\dot L_{\rm Edd}$ \cite{Maccarone2003}, and
that there is a hysteresis in the critical accretion rate value
depending on which direction the transition is going along
\cite{MaccaroneCoppi2003}.  Regardless of the exact details, a crucial
point for this paper is a phase-transition as a function of black hole
mass and accretion power.

We have previously suggested that the contribution of jets and
outflows on the spectral energy distribution (SED) of black holes can
be significant in supermassive as well as stellar mass black holes
\cite{FalckeMarkoff2000,MarkoffFalckeFender2001,Fender2001,YuanMarkoffFalcke2002} and that the jet contribution may in fact
  dominate the disk emission in a JDAF -- a jet-dominated accretion
  flow.  Jets are inherently broad-band, since they remain
  self-similar over many orders of magnitude in spatial scale and
  produce non-thermal particle distributions ranging over many orders
  of magnitude in energy. For this reason they should always be
  considered as potential contributors at every wavelength.  This
  concept of jet domination has now been empirically demonstrated for
  XRBs, where below $L_{\rm Edd}\approx 10^{-4}$ the kinetic energy
  output through radiatively inefficient jets (assumed to radiate only
  radio through IR) dominates the radiative output of the optically
  thin or thick disk (assumed to solely account for the X-ray
  emission; \citeNP{FenderGalloJonker2003}).  If the jet contributes
  to the X-rays as well, the jet domination may hold at even higher
  absolute luminosities.

The importance of jets to the emission of low-power accreting black
holes may hold the key to understanding the relationship between
stellar and galactic sized systems.  In the next section (Sect. 2) we
suggest how this concept can be used to provide a unified picture for
AGN as a function of mass and power for a range of sources that may be
operating at sub-Eddington accretion rates. This directly leads to a
prediction of radio/optical/X-ray scaling which we test on data from
several sources in Sect.~3.

\section{Low-Power Unification}
\subsection{A Scheme for Sub-Eddington Black Holes}

Our proposed scheme is based on three assumptions:

I) The accretion flow and disk form a coupled jet-disk system, with
jet and disk always being present in some form (``jet-disk
symbiosis'', see \citeNP{FalckeBiermann1995}).

II) Below a certain critical accretion rate, $\dot M_{\rm c}\simeq x
\times\dot M_{\rm edd}$ ($x\simeq0.01-0.1$), the inner part of the
accretion flow becomes radiatively inefficient
(e.g., \citeNP{EsinMcClintockNarayan1997}).

III) Below $\dot M_{\rm c}$,  or for face-on orientation
(relativistic beaming), the jet emission dominates the emission from
the accretion flow (e.g., \citeNP{YuanMarkoffFalcke2002}).

In short, the postulate is that {\it near-Eddington black holes are
disk-dominated and distinctly sub-Eddington black holes are
jet-dominated}.

Can we classify many of the various accreting systems we know of in
terms of this scheme, based on observational evidence?  Let us first
consider X-ray binaries where time scales are short enough that
individual sources can appear in a number of different states.  The
two most pronounced states are the high (soft) state, with a soft
power-law spectrum dominated by a thermal ``bump'', and the low (hard)
state characterized by a dominant hard power-law and weak-to-absent
thermal spectrum (e.g., \citeNP{Nowak1995}). The former is commonly
interpreted as multi-color blackbody emission from a standard thin
disk, while the latter is commonly attributed to an optically thin
accretion flow or corona. However, \citeN{MarkoffFalckeFender2001}
have suggested that the hard power law could also be attributed to
synchrotron emission from the jet in these systems. This is
strengthened by the finding of a tight non-linear correlation between
radio and X-ray luminosity in GX~339$-$4 \cite{CorbelNowakFender2003}
and other X-ray binaries in the low state
\cite{GalloFenderPooley2003}, which exactly fits the non-linear
predictions of the jet model (e.g., \citeNP{MarkoffNowakCorbel2003},
for GX~339$-$4). This correlation extends down into the quiescent
state, which is therefore now interpreted as an extremely low
luminosity hard state.  It has also been argued that some of the
ultra-luminous X-ray sources in nearby galaxies could be the beamed
equivalents of the well-known Galactic XRBs (``microblazars'';
\citeNP{MirabelRodriguez1999,KoerdingFalckeMarkoff2002}).

For AGN, the situation is more complicated since a large number of
source classes exist.  When considering higher luminosity sources with
strong disk signatures, the supermassive black hole equivalents to
soft state XRBs are FR~II radio galaxies, radio-loud quasars, and
blazars (with emission lines) among the radio loud objects. Within the
standard ``unified scheme'' these are mainly related through different
inclination angles.  On the radio quiet side, Seyfert galaxies,
radio-quiet quasars, and perhaps radio-intermediate quasars
\cite{MillerRawlingsSaunders1993,FalckeSherwoodPatnaik1996} are the
other analogs for high state XRBs. All of these AGN varieties show
direct or indirect evidence for a soft ultraviolet bump that can be
readily understood as emission from a standard accretion disk
\cite{SunMalkan1989}. This emission also provides ample photons to
produce the strong emission line regions seen in the optical spectra.

On the other hand, several low-power AGN classes seem to lack
evidence of a blue bump and strong emission lines, and are therefore
candidates for equivalents to the hard state XRBs. These are FR~I
radio galaxies, BL~Lacs and LINERs. The Galactic Center (Sgr A*; see
\citeNP{MeliaFalcke2001}) could also be in this category, but with its
faint and soft spectrum it is not clear what state in XRBs it would
correspond to. However, the almost-daily flares in Sgr A*
\cite{BaganoffBautzBrandt2001a} have a hard spectrum, so it may
therefore occasionally achieve a state analogous to the hard state in
XRBs.

In terms of radio power, FR~I radio galaxies form a smooth continuum
with FR~II radio galaxies, but are comparatively underluminous in
emission lines and lack a big blue bump
\cite{FalckeGopal-KrishnaBiermann1995,ZirbelBaum1995}. While FR~I
sources do seem to have optical cores, their fluxes scale tightly with
their radio flux \cite{ChiabergeCapettiCelotti1999}. This has been
used to argue for a synchrotron nature of these optical cores rather
then a thermal origin in the accretion disk. Interestingly, within the
standard unified scheme FR~I radio galaxies are coupled to BL~Lac
objects which are thought to be their relativistically beamed
versions. BL~Lacs -- by definition -- lack emission lines and there is
no population intermediate in inclination angle between FR~I and BL
Lacs which does show a blue bump or evidence for a standard optically
thick accretion disk.

Similarly, for low-luminosity AGN and LINERs, \citeN{Ho1999} argues
that their SED precludes the presence of a blue-bump and of a standard
accretion disk. On the other hand, radio observations of LINERs show a
strong jet presence \cite{FalckeNagarWilson2000,NagarWilsonFalcke2001}
and fits to individual objects indicate that the higher wavelengths
may also be dominated by jet emission \cite{YuanMarkoffFalcke2002b}. Some
of these LINERs are in big elliptical galaxies and may be the lower-luminosity
continuation of FR~I radio galaxies, while others sit in spiral
galaxies and may be somewhere in between Seyferts and our own Galactic
Center in terms of power.

Hence BL~Lacs, FR~Is, and LINERs are good candidates for sub-Eddington
and jet-dominated AGN.  Although this conclusion is already widely
accepted for BL~Lacs because of beaming arguments, and the case for FR
Is is strengthening, the proposal for LINERs remains highly debated. 

\begin{figure}[t]
\centerline{\hbox{\psfig{figure=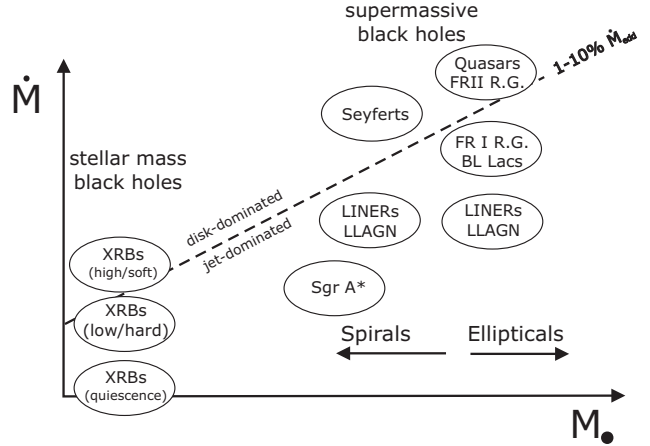,width=.47\textwidth,angle=0,bbllx=150pt,bblly=287pt,bburx=706pt,bbury=671pt}}}
\caption[]{
A proposed unification scheme for accreting black holes in the mass
and accretion rate plane. Above a few percent of the Eddington
accretion rate, the systems are proposed to be dominated by disk
emission, while below they are inherently dominated by jet emission
(RG=radio galaxy). Standard inclination-based unified schemes
(Antonucci 1993, Urry \& Padovani 1995) are still assumed to be valid
but are not explicitly shown here. Given a correlation between bulge
mass and black hole mass, the AGN with the most massive black holes
are supposed to reside in elliptical galaxies, while less massive
black holes are predominantly in spirals. This is, of course, not
applicable to XRBs.}\label{scheme}
\end{figure}

A sketch of the proposed scheme is shown in Fig.~\ref{scheme}. Note
that this is naturally very rough. In a number of cases the dividing
lines between individual classes may be blurred. Also, in
jet-dominated sources there may still be a sizeable disk contribution
and vice versa. In addition, as is commonly known, inclination effects
play an important role in unified schemes. For radio loud quasars, for
example, a small inclination to the line-of sight (i.e., in a blazar)
can lead to a significant jet contribution despite the fact that here
we classify these sources as intrinsically disk-dominated. This is in
contrast to BL~Lacs objects, which we consider as {\it intrinsically}
jet-dominated in addition to being beamed (with FR~I radio galaxies as
the parent population). This may have some analogy for XRBs, where
some Ultraluminous X-ray sources might be affected by beaming as well
\cite{KoerdingFalckeMarkoff2002}. In general the selection of BL~Lacs requires
significant care \cite{LandtPadovaniGiommi2002,MarchaBrowneImpey1996}
and the application of the scheme is not always straightforward
without good understanding of source properties and selection effects.

\nocite{Antonucci1993,UrryPadovani1995}

\subsection{Consequences and Tests of our Proposed Unification Scheme}

With such a scheme at hand, one wonders what the consequences are and
how they can be tested. First of all, if indeed black hole engines
make a qualitative transition with accretion power, a number of AGN
diagnostics have to be considered with even greater care. One example
is the ratio between radio and optical flux that is commonly used as a
radio-loudness parameter
\cite{KellermannSramekSchmidt1989,FalckeSherwoodPatnaik1996}. In most
interpretations it is supposed to represent the relative prominence of
jet and disk in a source. This has been particularly useful for
quasars, where one can well assume that the optical flux represents
disk emission. If, however, in sub-Eddington AGN both wavelengths are
coming from the jet, this parameter is physically no longer meaningful
as a jet-strength parameter and other factors have to be taken into
account.

This issue is particularly difficult when considering large samples of
AGN. Within each luminosity bin one can expect a range of black hole
masses to contribute and hence Eddington and sub-Eddington black holes
may be mixed if there are no well-sampled SEDs and spectra in radio,
optical, and X-rays. Moreover, mass itself can become a crucial
factor. This can in principle enhance scatter and spoil any possible
correlations or dichotomies.  On the other hand, if the SED of black
holes is jet-dominated, it may be possible to describe their evolution
with accretion power in a unified way. In the following we will now
concentrate on the expected scaling of radio, optical, and X-ray
emission from a jet-only model and compare it to data from samples of
sub-Eddington black holes.

\section{The X-ray/Radio Correlations}
\subsection{The Predicted Scaling}
\begin{figure}[t]
\centerline{\hbox{\psfig{figure=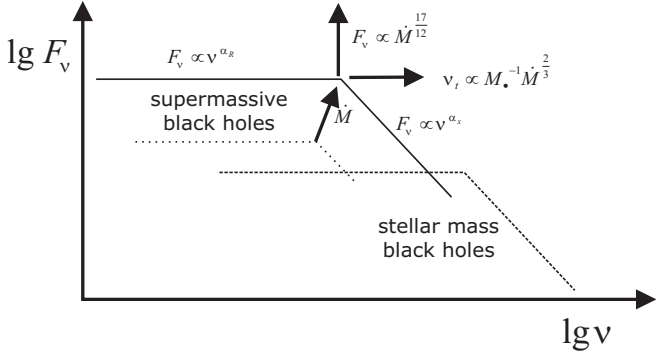,width=.47\textwidth,angle=0,bbllx=107pt,bblly=562pt,bburx=490pt,bbury=775pt}}}
\caption[]{A schematic jet spectrum and its theoretically expected scaling with mass and accretion rate. 
The spectrum has a flat-to-inverted, optically thick part below a turn-over
frequency $\nu_{\rm t}$ and a steep optically thin spectrum above. For
most sources the flat-to-inverted part of the spectrum will be in the
radio/infrared while the steep part will be in the optical and
X-rays. A change in the absolute accretion rate will shift the
spectrum along a diagonal line from the bottom left to the top
right. A change in mass will shift the spectrum horizontally
only. Lowering mass {\it and} accretion rate (e.g., by keeping the
accretion rate at a constant fraction of the Eddington accretion rate)
will shift the spectrum towards the bottom right, where stellar mass
black holes are found.}
\label{spectrum}
\end{figure}

Here we want to concentrate on the AGN core itself, leaving out the
extended emission.  In the simplest picture
\cite{BlandfordKonigl1979,FalckeBiermann1995}, the jet spectrum can be
naturally described by a flat-to-inverted radio spectrum up to a
turn-over frequency $\nu_{\rm t}$, which reveals an optically thin
power-law (see Fig.~\ref{spectrum}). The flat spectrum is the sum of
self-absorbed components along the jet, where higher frequencies
correspond to smaller regions closer to the black hole. The power law
results from optically thin emission from a power-law distribution of
electrons at the smallest scale in the jet where particle acceleration
exists. One can then roughly approximate the jet spetrum by a broken
power law normalized to a monochromatic luminosity (energy per time
and frequency) $L_{\rm t}$ at $\nu_{\rm t}$,

\begin{eqnarray}
L_{{\rm R}}&=&L_{\rm t}\left({\nu\over\nu_{\rm t}}\right)^{\alpha_{\rm R}}\quad\mbox{for}\;\nu\ll\nu_{\rm t} \quad\mbox{and}\label{rscale}\\
L_{{\rm X}}&=&L_{\rm t}\left({\nu\over\nu_{\rm t}}\right)^{\alpha_{\rm X}}\quad\mbox{for}\;\nu\gg\nu_{\rm t},
\end{eqnarray}
where $\alpha_{\rm R}\simeq0.15$ and $\alpha_{\rm X}\simeq-0.6$ are
the typical optically thick (radio) and optically thin (optical and
X-ray) spectral indices (e.g., \citeNP{MarkoffNowakCorbel2003}). This
spectrum can be mirrored to higher energies by inverse Compton
processes, leading, for example, to the characteristic 'camel's back'
SED of BL~Lacs in a $\nu L_\nu$ representation.

Scaling laws for this type of jet spectrum as a function of jet power
$Q_{\rm j}$ and mass have been described by \citeN{FalckeBiermann1995}
and \citeN{MarkoffNowakCorbel2003}. The main assumptions are that the
jet expands freely (conical shape), maintains an (arbitrary but fixed)
equipartition factor, and the distance of the first particle
acceleration zone, $z_{\rm acc}$, scales linearly with mass, e.g., is
always around some hundred to thousand $R_{\rm g}$. $R_{\rm
g}=GM_\bullet/c^2$ is the gravitational radius of the black hole.

As described in \citeN{FalckeBiermann1995} and
\citeN{MarkoffNowakCorbel2003}, it follows from simple analytic theory
that $L_{\rm t}\propto Q_{\rm j}^{17/12}$ and $\nu_{\rm t}\propto
Q_{\rm j}^{2/3}M_\bullet^{-1}$ and hence

\begin{eqnarray}\label{massscale}
L_{{\rm R}}&\propto& Q_{\rm j}^{{17\over12}-{2\over3}\alpha_{\rm R}}M_\bullet^{\alpha_{\rm R}}\left({\nu\over\nu_{\rm R}}\right)^{\alpha_{\rm R}}\quad\mbox{for}\;\nu\ll\nu_{\rm t}\quad \mbox{and}\\
L_{{\rm X}}&\propto& Q_{\rm j}^{{17\over12}-{2\over3}\alpha_{\rm X}}M_\bullet^{\alpha_{\rm X}}\left({\nu\over\nu_{\rm X}}\right)^{\alpha_{\rm X}}\quad\mbox{for}\;\nu\gg\nu_{\rm t},\label{freqscale}
\end{eqnarray}
where $\nu_{\rm R}$ and $\nu_{\rm X}$ are two fixed reference
frequencies. If we combine these equations we find the
expected radio/X-ray correlation
\begin{equation}
L_{\rm X} \propto L_{\rm R}^m M^{\alpha_{\rm X} - m \alpha_{\rm R}}
\label{eqRXScale}
\end{equation}
where
\begin{equation}
m = \frac{\frac{17}{12}-\frac{2}{3}
\alpha_{\rm X}}{\frac{17}{12}-\frac{2}{3} \alpha_{\rm R}}.
\end{equation}
Thus, to correct for the different masses of the objects we define an
equivalent optically thin (e.g., X-ray) luminosity
\begin{equation}
L'_{\rm X} = L_{\rm X} \left( \frac{\nu}{\nu_{\rm X}} \right)^{\alpha_{\rm X}}
\left(\frac{M}{6 M_{\odot}} \right)^{m \alpha_{\rm R}-\alpha_{\rm X}} \quad\mbox{for}\;\nu\gg\nu_{\rm t}.
\label{scaledX-rays}
\end{equation}
For the examples of $\alpha_{\rm R}\simeq0.15$ and $\alpha_{\rm
X}\simeq-0.6$ we get $m\simeq1.38$ and the mass correction factor is
predicted to go with $M^{0.81}$.

For relativistic steady jets, there will also be a dependency on the
Doppler factor ${\cal D}^{2-\alpha}$ (\citeNP{LindBlandford1985}).
However, since as a first-order approximation the monochromatic
luminosity at both frequencies is beamed by the same amount, the
correlation between $L_{{\rm R}}$ and $L_{{\rm X}}$ will only go as
$L_{{\rm R}}/L_{{\rm X}}\propto{\cal D}^{\alpha_{\rm X}-\alpha_{\rm
R}}$, i.e. less than linear for typical values. If there is a
significant velocity gradient along the jet, radio and X-rays could be
beamed by different amounts and the effect would become stronger.

Further parameters may affect the correlation. For example,
source-to-source variations in the equipartition factors or the
turn-over frequency $\nu_{\rm t}$ caused by different locations of the
first acceleration zone $z_{\rm acc}$ can lead to different
X-ray/radio ratios. However, since we have no good theoretical
understanding of such plasma parameters we have to accept this
uncertainty as a major source of scatter.

The scaling also only holds as long as a non-thermal power law is
produced in the optically thin regime and $\alpha_{\rm X}$ remains
roughly constant. This may not always be the case for sources that
approach quiescence, such as the Galactic Center in its non-flaring
state.

At least for an individual X-ray binary,
\citeN{MarkoffNowakCorbel2003} showed that this scaling with accretion rate can exactly
reproduce the tight non-linear radio/X-ray scaling of the X-ray binary
GX~339$-$4. Such a scaling has now been found to be fairly
representative for low-luminosity XRBs
\cite{GalloFenderPooley2003}. In the case of GX~339$-$4, the mass term
and Doppler factor were not included in the formula, since only one
source was considered. This is fine for XRBs, where the jet power and
accretion rate in one object changes over many orders of magnitude
within months and years. For AGN such changes take too long to be
discovered in individual objects and hence statistical samples have to
be used to cover a large range in instantaneous jet powers and
accretion rates.  In this case the mass becomes an important factor
for the thermal {\it and} non-thermal spectrum.

An important additional point concerns which wavelength to use in such
comparative studies of different source types and black hole masses.  In
the scaling law, we have been mainly comparing the optically thick
flux (mainly radio) to the optically thin flux (mainly X-rays). But,
it is important to know which wavelength belongs to which branch of
the jet SED in a certain type of source.  In essentially all sources
the {\it compact} radio emission is safely on the optically thick
branch of the jet core spectrum, however, since the turnover frequency
scales inversely with mass, the useful wavelength range over which one
can probe the optically thin branch of the SED may vary from one type
of source to another. We know, for example, that in BL~Lacs at least
the optical part of the SED belongs to the synchrotron branch. In some
cases this extends all the way into the X-rays in other cases,
however, X-rays may already be affected by the inverse Compton
components of the SED. Hence, optical flux measurements are a much
safer region for BL~Lacs (and FR~I radio galaxies for that matter) to
probe the optically thin part of the SED (which we still parameterize by
$\nu_{\rm X}$ and $L_{\rm X}$). On the contrary, X-ray binaries may
have a very high turn-over frequency, so that the optical flux may
still be on the optically thick branch (as discussed in
\citeNP{MarkoffFalckeFender2001}). Here, X-ray fluxes are 
the better choice, even though also here inverse Compton might
contribute. Whatever one chooses, a proper comparison requires one to
normalize the optically thin and optically thick fluxes to common
reference frequencies. This is done in
Eq.~(\ref{scaledX-rays}). Since we here use an X-ray frequency as
the common reference frequency for the normalized optically thin flux,
we stay with the term radio/X-ray correlation in the following, even
though it could for a number of sources equally well be a
radio/optical correlation.

\subsection{The Samples}
To test finally our hypothesis that the radio/X-ray correlation can be
traced from XRBs through LINERs, FR~Is, to BL~Lac objects, we use a
number of different samples from the literature where mass estimates,
radio and X-ray or optical fluxes have been published. For certain
types of sources (e.g. LLAGN) we are naturally limited by the small
number of well-defined samples that have been observed with the new
generation of X-ray telescopes.

For the XRBs we include the above mentioned multiple epochs of
GX~339$-$4 \cite{CorbelNowakFender2003}. We scaled the 8.6 GHz radio flux
to 5 GHz, assuming $(\alpha_{\rm R} = 0.15 )$.
 \citeN{HynesSteeghsCasares2003} give a mass for GX~339$-$4 around $6 M_{\odot}$. 
A distance of 4 kpc has been used to derive the luminosity
\cite{ZdziarskiPoutanenMikolajewska1998}. We note that other methods may give 
somewhat different distances (e.g., \citeNP{Maccarone2003}) and that
the mass is a strict lower limit. Nevertheless, the correlation for
GX~339$-$4 seems to be representative for a large number of XRBs in
the hard state \cite{GalloFenderPooley2003}.

As the lowest luminosity supermassive black hole, we included Sgr
$A^*$.  The 5 GHz radio flux was taken from the average spectrum in 
\citeN{MeliaFalcke2001}. The X-ray luminosity in the quiet and the flaring
state were taken from \cite{BaganoffBautzBrandt2001a}, which we
scaled with the given photon indices to a 3-9 keV luminosity.  The
black hole mass is taken to be $3\times10^6 M_{\odot}$
\cite{SchodelOttGenzel2002} and the distance of 8 kpc has been used.

For the LINERS we included the Chandra sample of
\citeN{TerashimaWilson2003}.  They selected 14 objects with radio
cores from the Low-Luminosity AGN (LLAGN) sample of
\citeN{NagarFalckeWilson2000}, with a flat or inverted radio core
$(\alpha_{\rm R} \geq -0.3 )$. \citeN{NagarFalckeWilson2000} selected
their sources from the \citeN{HoFilippenkoSargent1995} sample (a
magnitude limited sample) according to preliminary spectral
classification as LINER or as transitional object.  To compare the
X-ray luminosity with GX~339$-$4, we scaled the 2-10 keV X-ray
luminosity to a 3-9 keV luminosity assuming a power law index of
$\alpha_{\rm X}=-0.6$ for all objects.

For the FR~Is, we took the radio and HST data given in
\citeN{ChiabergeCapettiCelotti1999} who selected their sample from the
3CR catalogue \cite{SpinradMarrAguilar1985} which have been
morphologically identified as FR~I radio sources. The 33 sources form
a complete, flux limited sample.  The optical cores have been
extrapolated to a corresponding X-ray luminosity using
Eq.~(\ref{freqscale}) under the assumption that the synchrotron power
law has a spectral index of $\alpha_{\rm X}=-0.6$. We did not use
actual X-ray data, as the HST observations had higher resolution and
within the jet model for FR~I and BL~Lacs some of these high-mass
sources could have their synchrotron cut-off already below the X-ray
band, such that X-rays could be dominated by synchrotron self-Compton.

For the BL~Lacs we took X-ray (XBLs) and radio selected (RBLs) BL~Lacs
from \citeN{SambrunaMaraschiUrry1996}. These originate from two
complete samples: the Einstein Observatory Extended Medium-Sensitivity
Survey (EMSS) XBL sample \cite{MorrisStockeGioia1991} and the 1 Jy RBL
sample \cite{StickelFriedKuehr1991}.  Similar to FR~Is we calculate
the corresponding monochromatic X-ray luminosity from the optical data
assuming $\alpha_{\rm X}=-0.6$. Since BL~Lacs are thought to be
strongly affected by beaming, we corrected the radio and the
equivalent X-ray luminosity for Doppler boosting, assuming an average
Doppler factor of ${\cal D}\simeq7$
\cite{GhiselliniPadovaniCelotti1993}. 
As mentioned above, for the X-ray/radio correlation the Doppler factor
largely cancels out and enters less than linearly. Of course, the
position {\it along} the correlation will be affected more strongly.
For all source populations other than the BL~Lacs, we assume a Doppler
factor around unity.

For all sources we calculated the radio luminosity from the 5 GHz flux
density.  The distances of the sources were derived from the redshift
with $H_0 = 75$ km/s/Mpc. We selected from these samples all sources,
where we found black hole mass estimates in the literature or by using
the bulge velocity and the bulge/black hole mass relation from
\citeN{MerrittFerrarese2001}. Central velocity dispersion values were
taken from \citeN{PrugnielZasovBusarello1998} and its update in the
`Hypercat' database or from \citeN{WooUrry2002}.  The black hole
masses and fluxes are tabulated in Table 1.

\begin{table*}
          \label{tableAGN}
\setlength{\extrarowheight}{1pt}
\begin{center}
\begin{tabular}{p{0.12\linewidth}cccccc}
 Type/Name & Distance & M$_{\mbox{BH}}$ & $\mbox{F}_{\mbox{5GHz}}$ & $\mbox{F}_{\mbox{2-10keV}} $ & $\mbox{L}_{\mbox{5GHz}} $& $\mbox{L'}_{\mbox{3-9keV}} $\\
  &  [Mpc] & $[M_\odot]$ &  [mJy] & $\left[\mbox{erg}/\mbox{s}\right]$ & $\left[\mbox{erg}/ \mbox{s}\right] $& $\left[\mbox{erg}/ \mbox{s}\right]$\\
             \hline\hline
 SGR $A^*$&   &  &  &  &  & \\
\hline
Quiet& $0.008$ & $3.\times{10}^6$ & $600.$ & $2.2\times10^{33}$ & $2.3\times{10}^{32} $ & $ 6.3\times{10}^{37}$ \\
Flare & $0.008$ & $3.\times{10}^6$ & $600.$ & $1.0\times10^{35}$ & $2.3\times{10}^{32} $ & $ 2.87\times{10}^{39}$ \\
\hline 
 LLAGN &   &  &  &  &  & \\
             \hline
NGC2787 & $13.3$ & $1.7\times{10}^8$ & & $2.5\times{10}^{-14}$ & $1.66\times{10}^{37}$ & $ 3.93\times{10}^{44}$ \\
NGC3147 & $40.9$ & $6.58\times{10}^8$ & & $3.7\times{10}^{-12}$ & $1.02\times{10}^{38}$ & $ 1.65\times{10}^{48}$ \\
NGC3169 & $19.7$ & $6.21\times{10}^7$ & & $2.45\times{10}^{-12}$ & $1.55\times{10}^{37}$ & $ 3.76\times{10}^{46}$ \\
NGC3226 & $23.4$ & $1.39\times{10}^8$ & & $7.6\times{10}^{-13}$ & $1.58\times{10}^{37}$ & $ 3.15\times{10}^{46}$ \\
NGC4143 & $17.$ & $3.1\times{10}^8$ & & $3.1\times{10}^{-13}$ & $1.45\times{10}^{37}$ & $ 1.3\times{10}^{46}$ \\
NGC4278 & $9.7$ & $4.5\times{10}^8$ & & $8.1\times{10}^{-13}$ & $8.13\times{10}^{37}$ & $ 1.49\times{10}^{46}$ \\
NGC4548 & $16.8$ & $1.83\times{10}^7$ & & $1.6\times{10}^{-13}$ & $2.04\times{10}^{36}$ & $ 6.66\times{10}^{44}$ \\
NGC4565 & $9.7$ & $2.15\times{10}^7$ & & $3.2\times{10}^{-13}$ & $1.41\times{10}^{36}$ & $ 5.07\times{10}^{44}$ \\
NGC6500 & $39.7$ & $1.15\times{10}^8$ & & $3.\times{10}^{-14}$ & $7.94\times{10}^{38}$ & $ 3.08\times{10}^{45}$ \\
\hline 
 FR~I  &   &  &  &  &  & \\
             \hline
UGC00595 & $181.$ & $2.31\times{10}^8$ & $93.$ & $5.8\times{10}^{-18}$ & $ 1.82\times{10}^{40}$ & $ 5.84\times{10}^{48}$
 \\
NGC0383 & $67.8$ & $5.11\times{10}^8$ & $92.$ & $1.5\times{10}^{-17}$ & $ 2.53\times{10}^{39}$ & $ 4.02\times{10}^{48}$
\\
UGC01841 & $86.4$ & $1.78\times{10}^9$ & $182.$ & $4.93\times{10}^{-17}$ & $ 8.13\times{10}^{39}$ & $ 5.85\times{10}^{49
}$ \\
NGC1218 & $116.$ & $5.45\times{10}^8$ & $964.$ & $2.38\times{10}^{-16}$ & $ 7.75\times{10}^{40}$ & $ 1.96\times{10}^{50}
$ \\
NGC1275 & $70.7$ & $4.4\times{10}^8$ & $42400.$ & $1.5\times{10}^{-15}$ & $ 1.27\times{10}^{42}$ & $ 3.86\times{10}^{50}
$ \\
NGC3862 & $82.8$ & $4.9\times{10}^8$ & $200.$ & $1.14\times{10}^{-16}$ & $ 8.2\times{10}^{39}$ & $ 4.39\times{10}^{49}$
\\
NGC4261 & $29.6$ & $1.19\times{10}^9$ & $308.$ & $5.1\times{10}^{-18}$ & $ 1.62\times{10}^{39}$ & $ 5.16\times{10}^{47}$
 \\
NGC4374 & $14.8$ & $8.11\times{10}^8$ & $180.$ & $5.9\times{10}^{-17}$ & $ 2.36\times{10}^{38}$ & $ 1.09\times{10}^{48}$
 \\
NGC4486 & $14.8$ & $1.71\times{10}^9$ & $4000.$ & $3.9\times{10}^{-16}$ & $ 5.24\times{10}^{39}$ & $ 1.32\times{10}^{49}
$ \\
NGC5532 & $95.3$ & $8.67\times{10}^8$ & $77.$ & $3.4\times{10}^{-18}$ & $ 4.18\times{10}^{39}$ & $ 2.75\times{10}^{48}$
\\
UGC09799 & $138.$ & $2.48\times{10}^8$ & $391.$ & $9.6\times{10}^{-18}$ & $ 4.45\times{10}^{40}$ & $ 5.92\times{10}^{48}
$ \\
NGC6166 & $122.$ & $1.06\times{10}^9$ & $105.$ & $1.\times{10}^{-17}$ & $ 9.35\times{10}^{39}$ & $ 1.56\times{10}^{49}$ \\
NGC7236 & $105.$ & $1.22\times{10}^8$ & $2.$ & $9.1\times{10}^{-19}$ & $ 1.33\times{10}^{38}$ & $ 1.85\times{10}^{47}$ \\
UGC12064 & $72.7$ & $4.05\times{10}^8$ & $37.$ & $1.8\times{10}^{-17}$ & $ 1.17\times{10}^{39}$ & $ 4.59\times{10}^{48}$
 \\
NGC7720 & $121.$ & $1.22\times{10}^9$ & $270.$ & $1.9\times{10}^{-17}$ & $ 2.37\times{10}^{40}$ & $ 3.29\times{10}^{49}$
 \\
\hline 
 XBLs &   &  &  &  &  & \\
             \hline
0158+001 & $1270.$ & $1.13\times{10}^8$ & $11.3$ & $0.047$ & $1.1\times{10}^{41}$ & $ 8.93\times{10}^{50}$ \\
0257+342 & $1040.$ & $5.36\times{10}^8$ & $10.$ & $0.25$ & $6.49\times{10}^{40}$ & $ 1.12\times{10}^{52}$ \\
0317+183 & $792.$ & $8.12\times{10}^7$ & $17.$ & $0.36$ & $6.39\times{10}^{40}$ & $ 2.03\times{10}^{51}$ \\
0419+194 & $2260.$ & $4.73\times{10}^8$ & $8.$ & $0.09$ & $2.44\times{10}^{41}$ & $ 1.71\times{10}^{52}$ \\
0607+710 & $1130.$ & $5.27\times{10}^8$ & $18.2$ & $0.09$ & $1.39\times{10}^{41}$ & $ 4.68\times{10}^{51}$ \\
0737+744 & $1350.$ & $1.16\times{10}^9$ & $24.$ & $0.64$ & $2.6\times{10}^{41}$ & $ 8.92\times{10}^{52}$ \\
0922+745 & $2860.$ & $7.12\times{10}^9$ & $3.3$ & $0.044$ & $1.62\times{10}^{41}$ & $ 1.2\times{10}^{53}$ \\
1207+394 & $2750.$ & $1.78\times{10}^9$ & $5.8$ & $0.1$ & $2.63\times{10}^{41}$ & $ 8.22\times{10}^{52}$ \\
1221+245 & $914.$ & $8.33\times{10}^7$ & $26.4$ & $0.42$ & $1.32\times{10}^{41}$ & $ 3.23\times{10}^{51}$ \\
1229+643 & $680.$ & $4.17\times{10}^9$ & $42.$ & $0.55$ & $1.16\times{10}^{41}$ & $ 5.51\times{10}^{52}$ \\
1407+595 & $2180.$ & $3.08\times{10}^9$ & $16.5$ & $0.07$ & $4.68\times{10}^{41}$ & $ 5.62\times{10}^{52}$ \\
1534+014 & $1330.$ & $8.01\times{10}^8$ & $34.$ & $0.15$ & $3.61\times{10}^{41}$ & $ 1.52\times{10}^{52}$ \\
1757+703 & $1770.$ & $6.92\times{10}^8$ & $7.2$ & $0.18$ & $1.34\times{10}^{41}$ & $ 2.85\times{10}^{52}$ \\
2143+070 & $998.$ & $3.13\times{10}^8$ & $50.$ & $0.32$ & $2.98\times{10}^{41}$ & $ 8.53\times{10}^{51}$ \\
\hline 
 RBLs  &   &  &  &  &  & \\
             \hline
1418+546 & $629.$ & $1.46\times{10}^9$ & $1220.$ & $2.72$ & $2.89\times{10}^{42}$ & $ 1.\times{10}^{53}$ \\
1807+698 & $206.$ & $2.67\times{10}^{10}$ & $1710.$ & $7.85$ & $4.36\times{10}^{41} $ & $ 3.24\times{10}^{53}$ \\
2005-489 & $289.$ & $1.48\times{10}^9$ & $1210.$ & $9.85$ & $6.03\times{10}^{41}$ & $ 7.71\times{10}^{52}$ \\
2200+420 & $280.$ & $1.71\times{10}^8$ & $2140.$ & $8.65$ & $1.01\times{10}^{42}$ & $ 1.12\times{10}^{52}$ \\
2254+074 & $792.$ & $4.82\times{10}^8$ & $560.$ & $0.6$ & $2.1\times{10}^{42}$ & $ 1.43\times{10}^{52}$ \\
           \hline
          \end{tabular}
\end{center}

\caption[]{Sources used in this paper. Column 1 lists the names of the sources and column 2 gives
the distance used to derive the luminosities from the fluxes. The black hole
mass was calculated from the velocity dispersion relation 
\citeN{MerrittFerrarese2001}. Column 4 and 5 give the measured radio
and X-ray fluxes. For the LLAGN sample we only list the radio
luminosity as directly taken from the original paper.  The last two
columns give the radio luminosity and the equivalent X-ray luminosity
as described in Eq.~(\ref{scaledX-rays}). This luminosity has also
been corrected for the different observed energy bands assuming a
photon index of 1.6.}
\end{table*}

\subsection{Results}
\begin{figure}[t]
\centerline{\psfig{figure=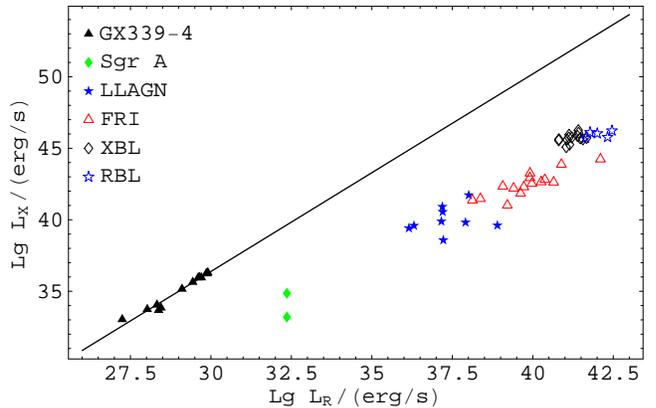,width=.47\textwidth,angle=0}}
\caption[]{Radio/X-ray correlation for XRBs with our AGN sample. We only extrapolate the
optical measurements of some AGN (FR~I radio galaxies) to a
corresponding monochromatic X-ray luminosity without a mass
correction.  For Sgr A* we show the quiescent and the flare
spectrum. The solid line is the analytically predicted non-linear
radio/X-ray correlation from the jet model, normalized for
GX339-4. The supermassive black holes fall below the extrapolation
from the X-ray binaries.  }\label{xr-correlation}
\end{figure}

\begin{figure}[t]
\centerline{\psfig{figure=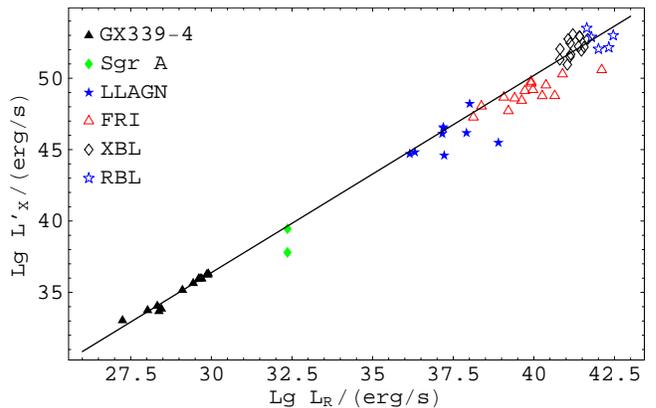,width=.47\textwidth,angle=0}}
\caption[]{
The same as Fig.~\ref{xr-correlation} but for an equivalent X-ray
luminosity, $L'_{\rm X}$, which has been individually corrected for
the mass factor and scaled to the value the X-ray luminosity would
have for a central black hole of only 6 $M_{\odot}$, as in GX339-4
(see Eq.~\ref{scaledX-rays}). Corrections for Doppler factors have not
been applied. }\label{xprimer-nodop}
\end{figure}

\begin{figure}[t]
\centerline{\psfig{figure=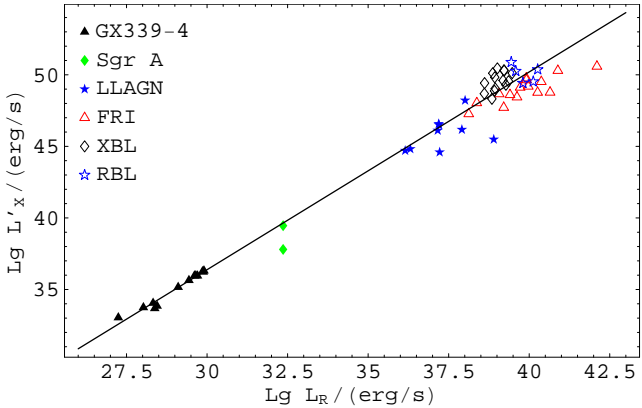,width=.47\textwidth,angle=0}}
\caption[]{
The same as Fig.~\ref{xprimer-nodop} but the radio and X-ray
luminosities of BL~Lac objects have been corrected for Doppler
boosting. As discussed in the text, this mainly moves BL~Lacs along
the correlation and they now occupy the same region as FR~Is -- their
parent population within the inclination-based unified scheme.
}\label{xprimer-correlation}
\end{figure}

\begin{figure}[t]
\centerline{\hbox{\psfig{figure=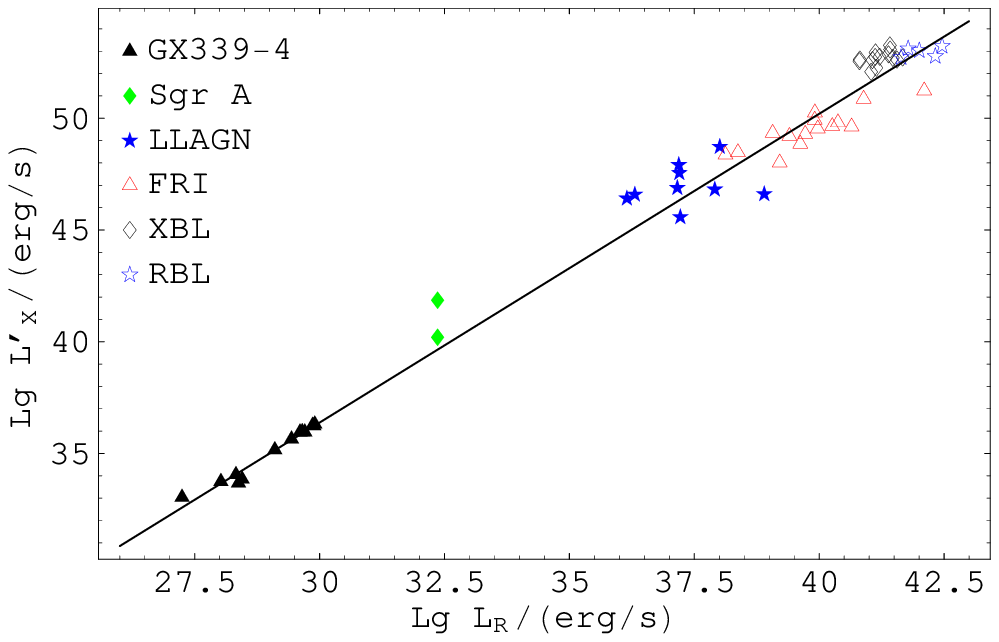,width=.47\textwidth,angle=0}}}
\caption[]{
Radio/X-ray correlation for XRBs and AGN, where the X-ray flux of all
AGN has been increased by a constant value of $10^7$, corresponding to
an average AGN mass of $3\times10^9 M_\odot$. }\label{scaledE7}
\end{figure}

In Figs.~\ref{xr-correlation} to \ref{scaledE7} we show the radio and
X-ray luminosities of the sources discussed above with various
correction factors applied. Figure \ref{xr-correlation} shows the
uncorrected data, with only optical luminosities extrapolated to
corresponding X-ray luminosities. Clearly, the AGN fall well below the
extrapolation of the radio/X-ray correlation of X-ray binaries. In
other terms: by simply increasing the accretion rate in an X-ray
binary one will never obtain the SED of an AGN.

In Fig.~\ref{xprimer-nodop} we have included in the correlation the
analytically predicted mass scaling (Eq.~\ref{scaledX-rays}) but not
yet the correction of the Doppler factor for BL~Lacs.  Surprisingly,
with this simple scaling, all the source populations seem to be scaled
by just the right amount to fall more or less on the predicted scaling
with power from the XRBs with a relatively low scatter.  This means
that in the parameter space of X-ray luminosity, radio luminosity, and
black hole mass, sub-Eddington black holes form a fundamental plane.
It also suggests, that the theoretically motivated and predicted
scaling seems to hold for stellar mass as well as supermassive black
holes.  We point out that two of the outliers (NGC6500 and NGC1275)
are known from high-resolution VLBI observations to have radio cores
that are significantly affected by extended emission
\cite{FalckeNagarWilson2000,WalkerDhawanRomney2000} and hence appear
too bright in the radio. The same may be true to some degree for FR~Is
in general, but should be negligible for BL Lacs.

Figure \ref{xprimer-correlation} demonstrates the effect of a Doppler
factor correction. As discussed in the previous section, the
X-ray/radio-ratio is rather insensitive to the Doppler factor and
sources will mainly move along the correlation. In our case the
BL~Lacs are pushed from the upper end of the correlation into the
regime where FR~I radio galaxies lie. Given that BL~Lacs are supposed
to be the beamed population of FR~I radio galaxies within the
inclination-based unified scheme, this seems to be an appropriate
correction and provides further support for that scheme.

Finally, in Fig.~\ref{scaledE7} we show the radio and X-ray
luminosities, where the X-ray flux has been corrected by a constant
factor $10^7$, thus ignoring the individual mass estimates. With this
factor the radio/X-ray correlation can also be continued to
AGN. Scaling by $10^7$ is identical to assuming a constant black hole
mass of $\simeq3\times10^9 M_{\odot}$ for all objects. The black hole
mass of FR~I Radio Galaxies and BL~Lac objects scatter around this
value. LLAGN have an average mass of somewhat less than $10^9
M_{\odot}$, thus in comparison with Fig.~\ref{xprimer-correlation},
the LLAGN have higher X-ray fluxes. The Galactic black hole (Sgr
A$^*$) has a mass of only $3\times10^6 M_{\odot}$ so the X-ray flux is
increased too much and the X-ray flare state -- which may in fact
contain the here crucial non-thermal power law -- lies above the
extrapolation.

A better distinction of the mass effects might be possible with the
inclusion of more low mass AGN. Another conclusion is that, for
example, a linear dependency of the X-ray/radio-ratio with mass would
not be appropriate and over-correct the data.

\section{Conclusion and Discussion}
We have suggested that black holes operating at sub-Eddington
accretion rates make a transition to a radiative inefficient state,
where most of the emission is largely dominated by the non-thermal
emission of a jet (``JDAFs''). In this picture the radiative output of
sub-Eddington black holes is non-thermally dominated, while
near-Eddington black holes are thermally dominated. This scheme allows
one to unify the radiative properties of black holes over a large
range of accretion powers. At sub-Eddington accretion rates, the
scaling between radio and optical or X-ray cores is then predicted to
follow the scaling laws outlined in
\citeN{FalckeBiermann1995} and \citeN{MarkoffNowakCorbel2003}. This
requires taking the black hole mass into account.

Near-Eddington black holes are presumably found in quasars, luminous
Seyfert galaxies, and soft-state X-ray binaries which are considered
to be in the high state.  As pointed out elsewhere
\cite{PoundsDoneOsborne1995,MaccaroneGalloFender2003} Narrow-Line
Seyfert 1s may also be related to the very high state of X-Ray
binaries.

On the other hand, candidates for sub-Eddington black holes are XRBs
in the low-hard state, Sgr A*, LINERs, FR~I radio galaxies, and BL~Lac
objects. In terms of beaming and inclination-based unified schemes,
which we do not explicitly discuss but consider valid, it may be worth
pointing out that ultra-luminous X-ray sources might be low-mass
analogs to BL~Lacs and blazars
\cite{KoerdingFalckeMarkoff2002}.

Using various samples of sub-Eddington black holes, we are able to
show that all these different types of sources seem to fall near the
predicted radio/X-ray correlation, if the scaling with black hole mass
is taken into account.

The crucial underlying assumption is that all these latter sources are
intrinsically jet-dominated and have essentially the same SED in
common: a flat, optically thick radio spectrum and an optically thin
power law beyond a turn-over frequency.  Shape and scaling of the SED
needed to explain the radio/X-ray correlation is just what one expects
in a pure jet model and supports the notion of jet-dominated accretion
flows (``JDAF''). On the other hand, some form of radiative
inefficient accretion flows/corona is also clearly needed for this
picture to work, since there is always a need for a power and matter
source for the outflow. It may be possible to adapt the scheme fo a
situation where the X-ray emission is dominated by emission from
optically thin accretion flows, if their X-ray flux follows a similar
non-linear scaling as predicted in the jet case.

An interesting corollary for jets is that, in order to obtain the
scaling with mass, one has to assume that the region of the onset of
particle acceleration in the jet -- producing the optically thin power
law -- is always around a fixed location in mass-scaled units ($\sim
100-1000 R_{\rm g}$).

With the large range of black hole powers and masses discovered the
proposed picture may warrant further investigation and detailed
tests. If solidified and further evolved it may help to predict the
luminosity evolution of black holes at various wavebands over many
orders of magnitude. Interestingly, two other papers have recently
appeared that come to very similar conclusion in terms of the expected
scaling \cite{HeinzSunyaev2003} and its application to black holes of
different masses \cite{MerloniHeinzMatteo2003}.

\begin{acknowledgements}
We are thankful to Tom Maccarone, Rob Fender, and an anonymous referee
for helpful comments that improved the masnucript.  S.M. is supported
by an NSF Astronomy and Astrophysics Postdoctoral Fellowship, under
award AST-0201597.
\end{acknowledgements}


\end{document}